\documentclass[sigconf,screen,nonacm]{acmart}

\setcopyright{none}
\renewcommand\footnotetextcopyrightpermission[1]{}

\acmConference[DAC '26]{Design Automation Conference}{July 26--29, 2026}{Long Beach, CA, USA}

\usepackage{amsmath,amssymb,amsfonts}
\usepackage{graphicx}
\usepackage{textcomp}
\usepackage{xcolor}
\usepackage{algorithm}           
\usepackage{algpseudocode}       
\usepackage{subcaption}
\usepackage{booktabs}
\usepackage{multirow}
\usepackage{tabularx}
\usepackage{siunitx}
\usepackage{tikz}
\newcommand*\circled[1]{\tikz[baseline=(char.base)]{
    \node[shape=circle,fill=black,text=white,draw,inner sep=1.2pt] (char) {#1};}}

\AtBeginDocument{%
  }

\usepackage{caption}
\begin{document}

\title{ZK-Tracer: A High-Performance Heterogeneous Accelerator for Zero-Knowledge VM Trace Generation}
\titlenote{This paper has been accepted by DAC 2026 and will appear in the proceedings.}

\author{\large
    Jieran Cui\textsuperscript{\rm 1,2},
    Zhengkai Wen\textsuperscript{\rm 1,2},
    Haowen Fang\textsuperscript{\rm 2},
    Yinan Zhu\textsuperscript{\rm 1,2},
    Jia Xiong\textsuperscript{\rm 1,2},\\
    Cheng Ni\textsuperscript{\rm 1,2},
    Mingchi Zhang\textsuperscript{\rm 1,2},
    Nan Guan\textsuperscript{\rm 3},
    Xi Wang\textsuperscript{\rm $\ast$1,2} \\
    \textsuperscript{\rm 1}School of Integrated Circuits, Southeast University, China \\
    \textsuperscript{\rm 2}National Center of Technology Innovation for EDA, China \\
    \textsuperscript{\rm 3}Department of Computer Science, City University of Hong Kong, Hong Kong \\
    [-0.6ex]
    \textsuperscript{$\ast$} Corresponding Author: xi.wang@seu.edu.cn
}
\renewcommand{\shortauthors}{Jieran Cui et al.}

\begin{abstract}
Zero-knowledge virtual machines (zkVMs) are a key technology for driving the large-scale adoption of zero-knowledge proofs (ZKP), but their performance bottlenecks severely limit their practicality. While current hardware acceleration research has exclusively focused on backend proving, we identify that the frontend execution and trace generation phase is rapidly emerging as the new system bottleneck. To address this challenge, we propose ZK-Tracer, \textbf{the first hardware accelerator architecture} specifically designed for the zkVM frontend. ZK-Tracer features a novel heterogeneous design comprising a Main Trace Unit and parallel Permutation Trace Units. It exposes a fine-grained interface to the host software through a lightweight instruction set extension, enabling efficient task offloading. Our ASIC implementation results demonstrate that ZK-Tracer achieves up to  \textbf{1829$\times$} speedup in trace generation over a high-performance multi-core CPU. When integrated with existing backend proving accelerators, it delivers a remarkable  \textbf{963$\times$} end-to-end performance improvement for the entire ZKP system.

\end{abstract}


\maketitle

\section{Introduction}
Zero-Knowledge Proofs (ZKPs)~\cite{zkp-19} are the cornerstone of building the next generation of trustworthy digital systems. They enable a prover to convince a verifier of the validity of a statement without revealing any additional information. This revolutionary property has demonstrated immense potential in cutting-edge domains such as Ethereum scaling (e.g., ZK-Rollups)~\cite{xie2022zkbridge,kosba2016hawk,sasson2014zerocash}, privacy-preserving computation~\cite{danezis2013pinocchio,delignat2016cinderella}, and verifiable AI~\cite{zhang2020zero,chen2024zkml,feng2024zeno}, attracting widespread attention from both academia and industry.

\begin{figure}[t]
  \centering
  \includegraphics[width=1\linewidth, height=0.6\textheight, keepaspectratio]{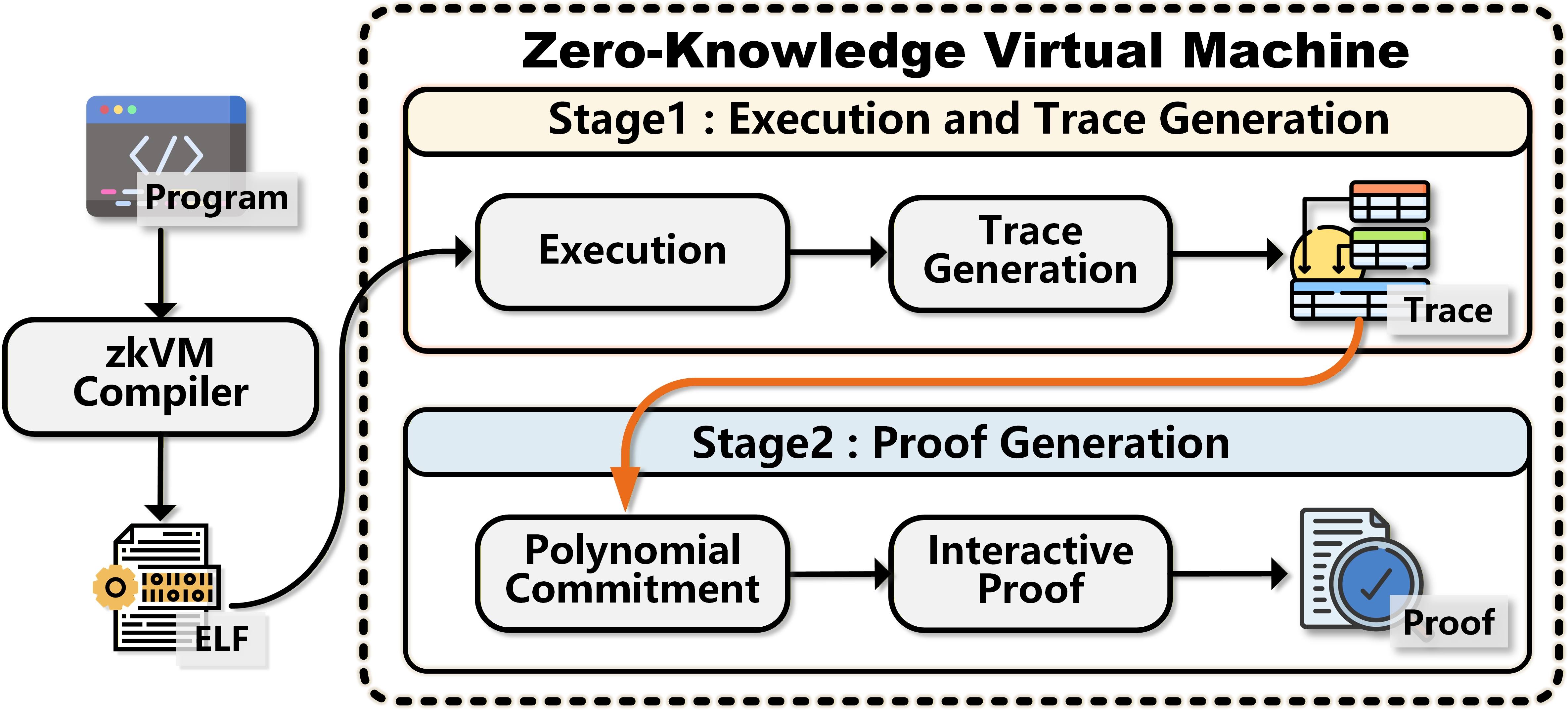}
  \caption{zkVM Workflow}
  \label{fig:zkVM_Workflow}
\end{figure}

However, traditional ZKP development demands profound cryptographic expertise to manually design arithmetic circuits, posing an extremely high barrier to entry. To overcome this hurdle, the Zero-Knowledge Virtual Machine (zkVM) has emerged. By allowing developers to program in high-level languages like Rust, zkVMs significantly lower the development barrier and are thus regarded as a key enabler for the large-scale adoption of ZKPs. Nevertheless, their performance has become the primary bottleneck hindering widespread application. Compared to native compiled execution, a zkVM can be hundreds of thousands to millions of times slower~\cite{zkVMs_slow}. This enormous performance gap renders them impractical for many real-world applications. A complete zkVM workflow, as illustrated in Figure \ref{fig:zkVM_Workflow},  consists of two main stages: \textbf{(1) Front-end Execution and Trace Generation}, which involves interpreting the program and recording its computational steps; and \textbf{(2) Back-end Proof Generation}, which uses the generated trace to produce the final proof via complex cryptographic algorithms, such as zk-STARK and zk-SNARK~\cite{oude2024systematic}.

Currently, hardware acceleration research in both academia and industry has exclusively focused on the computationally intensive back-end proving stage~\cite{acceleratingNTT-ASPLOS25,Nocap-MICRO24,zkSpeed-ISCA25,amaze-ICCAD24,zeno-ASPLOS24,sam-ICCAD23,myosotis-TCAD24,acclmt-DAC25,fully-TCAD24}. 
These efforts commonly treat the execution trace as a pre-existing, static input, thereby overlooking the overhead of its front-end generation. 
Our research, however, reveals that this ``back-end-centric'' optimization view is becoming outdated. 
Through an in-depth performance analysis of a state-of-the-art zkVM (Succinct SP1~\cite{SP1}), we found that for a suite of representative programs, the front-end execution and trace generation stage already accounts for 20\% to 30\% of the end-to-end proving time. 
According to Amdahl's Law, as back-end accelerators continue to improve, this unoptimized front-end is rapidly evolving into the dominant system performance bottleneck. 
For instance, even with a conservative 10$\times$ speedup in the back-end, the share of the front-end in the runtime would surge to over 80\%, severely capping any further end-to-end performance gains.

To address this emerging and critical challenge, this paper proposes \textbf{ZK-Tracer, the first hardware accelerator architecture specifically designed for zkVM front-end execution and trace generation}. Our work serves as a crucial complement to existing back-end proof accelerators, aiming to enable a full-stack, high-performance, end-to-end zkVM solution. 

The main contributions of this paper are as follows:

\begin{itemize}

    \item We design ZK-Tracer, a novel heterogeneous co-design architecture. It comprises a Main Trace Unit (MTU) for efficient program execution and Main trace generation, and Permutation Trace Units (PTUs) for parallel processing of the Permutation Trace. These units work efficiently through a custom Instruction Set Architecture (ISA).
    \item We introduce a suite of algorithm-hardware co-design innovations that exploit the mathematical properties and structural characteristics of ZKP traces. Key techniques include: (1) zero-overhead trace capture via non-intrusive snooping coupled with on-the-fly modular reduction; (2) a precomputation scheme slashing modular exponentiation complexity from $O(N \cdot k)$ to $O(k)$; and (3) a dedicated batch modular inversion unit achieving near-linear $O(N)$ complexity.
    
    \item We implemented ZK-Tracer on the TSMC 28nm process and conducted a comprehensive evaluation. Experimental results demonstrate that compared to a software baseline running on a high-performance x86 CPU, ZK-Tracer achieves up to a \textbf{1829$\times$ speedup} in trace generation. When integrated with a state-of-the-art prover~\cite{zkSpeed-ISCA25}, our work promises up to a \textbf{963$\times$ end-to-end speedup}, thereby delivering significant end-to-end acceleration for ZKP system.

\end{itemize}

\section{Motivation and Related Work}

\begin{figure}[t!]
  \centering
  \includegraphics[width=1\linewidth, height=0.6\textheight, keepaspectratio]{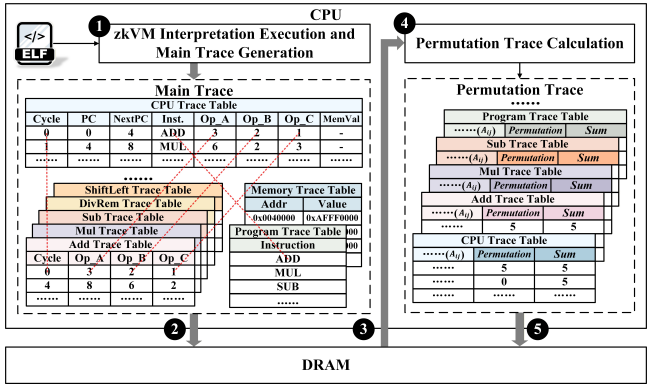}
  \caption{zkVM Trace Generation Flow}
  \label{fig:Trace_Structure}
\end{figure}

\begin{figure*}[t!]
  \centering 
  \begin{minipage}[b]{0.62\linewidth}
    \centering
    \includegraphics[width=\linewidth, height=0.3\textheight, keepaspectratio]{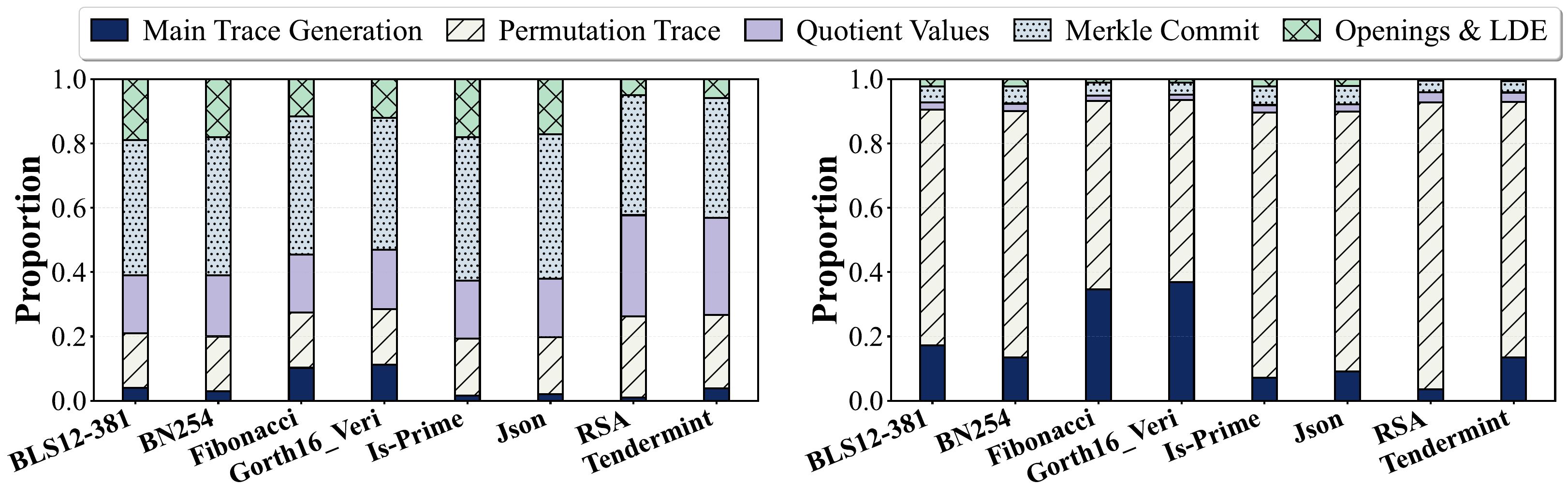}
    \captionof{figure}{Profiling before (left) and after backend acceleration (right)}
    \label{fig:zkVM_Runtime_Profiling}
  \end{minipage}%
  \hfill %
  \begin{minipage}[b]{0.37\linewidth}
    \centering
    \includegraphics[width=\linewidth, height=0.3\textheight, keepaspectratio]{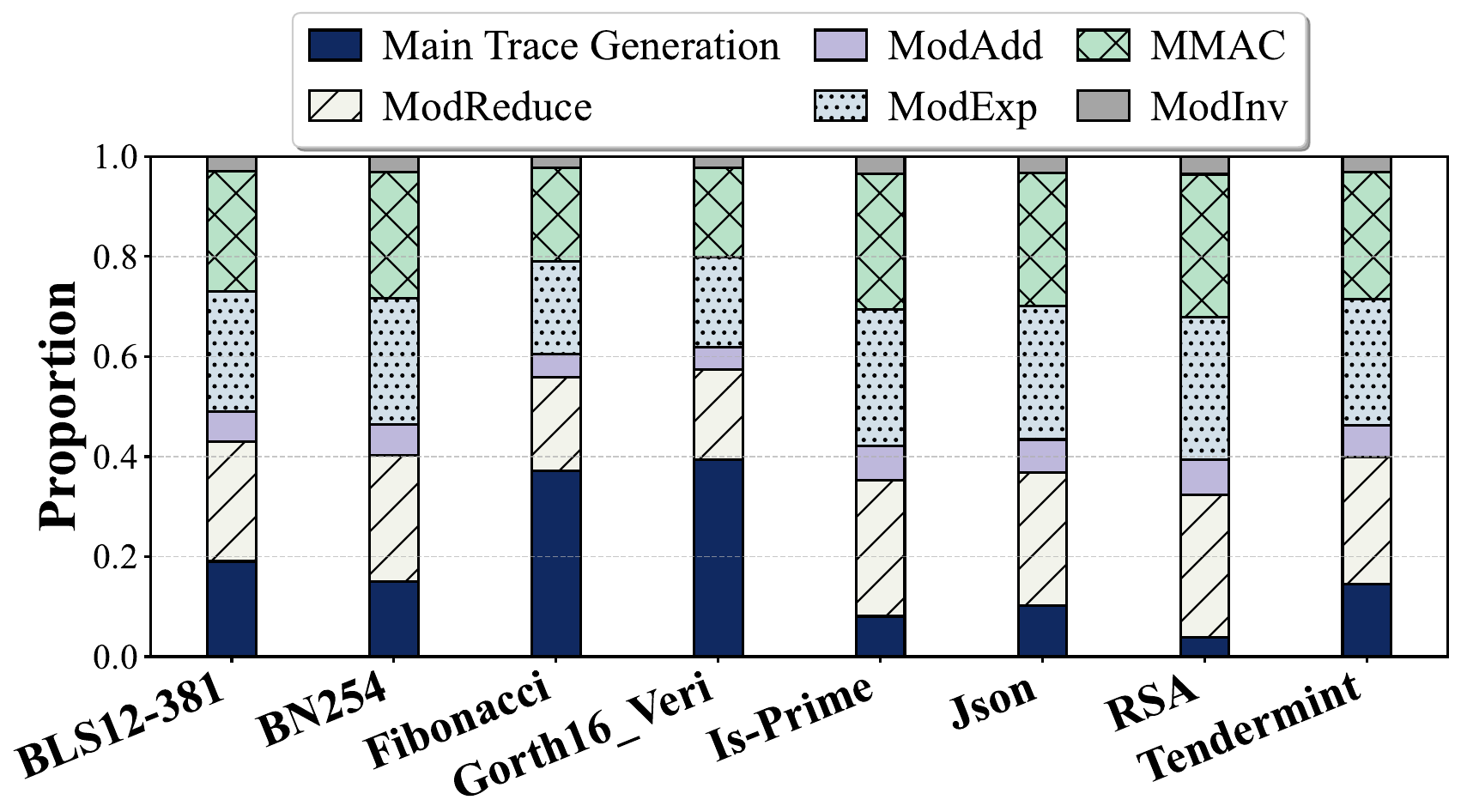}
    \captionof{figure}{Trace Generation Workload Analysis}
    \label{fig:Workload_Analysis}
  \end{minipage}
\end{figure*}

\subsection{ZKP Bottleneck-Shifting Phenomenon}
In recent years, hardware acceleration has emerged as a crucial approach to overcome the prohibitive computational overhead of Zero-Knowledge Proof (ZKP) systems. 
To date, research efforts from both academia and industry have focused exclusively on accelerating the \textbf{backend} proving phase. 
By designing dedicated hardware for compute-intensive primitives such as Multi-Scalar Multiplication (MSM)~\cite {gypsophila-DAC24, msmac-DAC24, cuzk-IACR23,hardcaml-FPGA24,falic-TC24,MSMaccelerating-ASPLOS24}, Number-Theoretic Transform (NTT)~\cite{pipezk-ISCA21,proteus-TVLSI24,gzkp-ASPLOS23,legozk-HPCA25,szkp-PACT24}, and hash computations~\cite {Nocap-MICRO24,unizk-ASPLOS25,zkSpeed-ISCA25,batchzk-ASPLOS25}, these works have reduced proof generation times by two to three orders of magnitude.
The very success of these ZKP backend accelerators, however, has fundamentally altered the performance landscape. As dictated by Amdahl's Law, the relative overhead of the \textbf{front-end} execution and trace generation stage has been dramatically magnified, transforming it into an emerging bottleneck that now constrains the end-to-end performance of the entire ZKP system.

To quantitatively illustrate this bottleneck-shifting phenomenon, we conducted fine-grained runtime profiling on a suite of representative programs (e.g., hashing, signature verification, and blockchain verification) using a typical zkVM system (SP1), with the results depicted in Figure~\ref{fig:zkVM_Runtime_Profiling}(left). 
In the current software-only implementation, the front-end execution and trace generation phase already accounts for 20\% to 30\% of the end-to-end proving time. 

Furthermore, by referencing performance reports from mainstream backend accelerators~\cite{pipezk-ISCA21,Nocap-MICRO24,legozk-HPCA25,Rezk-TCAS-I,unizk-ASPLOS25,batchzk-ASPLOS25,gzkp-ASPLOS23,zkSpeed-ISCA25}, we estimate that the overall average acceleration ratio of their backends is approximately 184$\times$.
By conservatively assuming that only 20\% of this backend acceleration is achieved, the runtime breakdown is shown in Figure~\ref{fig:zkVM_Runtime_Profiling}(right). 
The analysis reveals that the proportion of front-end overhead would surge dramatically to over \textbf{90\%}, becoming the overwhelmingly dominant bottleneck.

This analysis indicates that the substantial gains from backend accelerators are being significantly offset by the ever-growing front-end overhead, thereby limiting the overall end-to-end performance improvement. 
Consequently, hardware acceleration for zkVM front-end trace generation is no longer a ``nice-to-have'' optimization but an imperative. 
It is an essential step to break the performance ceiling of current ZKP systems and unleash the full potential of backend acceleration.

\subsection{Execution Trace in zkVM}

The execution trace is a core data structure in a zkVM, meticulously recording the machine state of a program execution in a tabular, chronological format. To reduce constraint complexity and enhance proving efficiency, modern high-performance zkVMs (e.g., RISC0~\cite{RISC0}, SP1~\cite{SP1}) adopt a multi-table architecture. Instead of consolidating all state into a single monolithic table, this architecture employs multiple specialized tables that operate in concert. The \textbf{CPU table} acts as the central hub, chronicling the instruction flow and state evolution. Concurrently, specialized tables, such as \textbf{ALU tables}, handle distinct operations like arithmetic and logic. \textbf{Memory and program tables} record the initial and final states of memory and code, respectively, to verify access consistency. To guarantee cross-table consistency, these systems widely employ a lookup argument based on logarithmic derivatives (LogUp)~\cite{SP1Whitepaper}. This mechanism unifies inter-table interactions and memory accesses into ``send/receive'' pairs, with \emph{multiset equality} maintained via permutation/lookup accumulators. 
Consequently, the execution trace inevitably comprises two components: the \textbf{Main Trace}, directly generated by the program's semantics, and the \textbf{Permutation Trace}, constructed for cross-table communication. The Permutation Trace is computed as formulated in Equation~\ref{eq:perm_trace_computation}, where $\gamma$ and $\beta$ are random challenges and $A_{ij}$ are values from the trace tables.
\begin{equation}
\text{Permutation}_i = \frac{1}{\gamma + \sum_j \beta^j A_{ij}} \ \ \ \ \ \ \ \text{Sum}_i = \sum_{k=1}^{i} \text{Permutation}_i
\label{eq:perm_trace_computation}
\end{equation}

\vspace{-6pt}
\subsection{Trace Generation Workload Analysis}

Figure \ref{fig:Trace_Structure} illustrates the typical workflow for generating a zkVM execution trace on a general-purpose CPU. The process begins with \circled{1} the interpretive execution of the Guest program to generate the main trace on-the-fly, which is then \circled{2} written to DRAM. After the entire main trace generation is complete, the process \circled{3} reads all of this data back from DRAM. Based on this data, it then \circled{4} calculates the permutation and accumulator columns required for cross-table communication, forming the permutation trace. Finally, \circled{5} this permutation trace is written back to memory for consumption by the subsequent proving phase. This process reveals that the root causes of the slow performance on a general-purpose CPU include:

\begin{itemize}
    \item \textbf{Interpretive Execution Overhead.}
    Executing the Instruction Set Architecture (ISA) of a guest program via software interpretation on a host CPU is inherently 1--2 orders of magnitude slower than native execution. 

    \item \textbf{Data Amplification and Memory Bottlenecks.}
    A single guest instruction generates multiple trace entries, leading to significant \textit{data amplification}. Critically, the multi-pass \textbf{write-read-write} access pattern exhibits poor temporal locality, causing cache thrashing and thus increasing latency and power consumption.

    \item \textbf{Limited Parallelism.}
    
 Its fundamental operations, including modular inversion and the sequential update of accumulator columns, expose little Data-Level Parallelism (DLP) or Instruction-Level Parallelism (ILP). As a consequence, key architectural mechanisms such as superscalar execution and out-of-order scheduling offer limited benefit, since the workload is dominated by long dependency chains with minimal independent instructions to exploit.
\end{itemize}

We further performed a workload decomposition of the operators involved in the trace generation phase, as illustrated in Figure \ref{fig:Workload_Analysis}. Aside from the main trace generation, the permutation trace calculation is predominantly composed of modular arithmetic. Key operations such as Modular Reduction, Modular Multiply-Accumulate (MMAC), and Modular Exponentiation each account for over 20\% of the workload, whereas Modular Addition is comparatively less frequent. Although Modular Inverse constitutes a small fraction of the total operations, its high latency per operation makes it a critical bottleneck, particularly in software implementations. 
Overall, the permutation trace calculation is characterized by its regular computation patterns, high degree of parallelism, and streaming-friendly dataflow. Based on this analysis, the zkVM execution trace generation task exhibits significant potential for hardware acceleration. By designing efficient memory access channels and dedicated parallel computation units, the core computational workload of trace generation can be shifted from inefficient software emulation to high-speed native hardware execution, thereby achieving substantial acceleration.

\section{Design and Philosophy}

\begin{figure*}[t!]
  \centering
  \includegraphics[width=\linewidth, height=0.55\textheight, keepaspectratio]{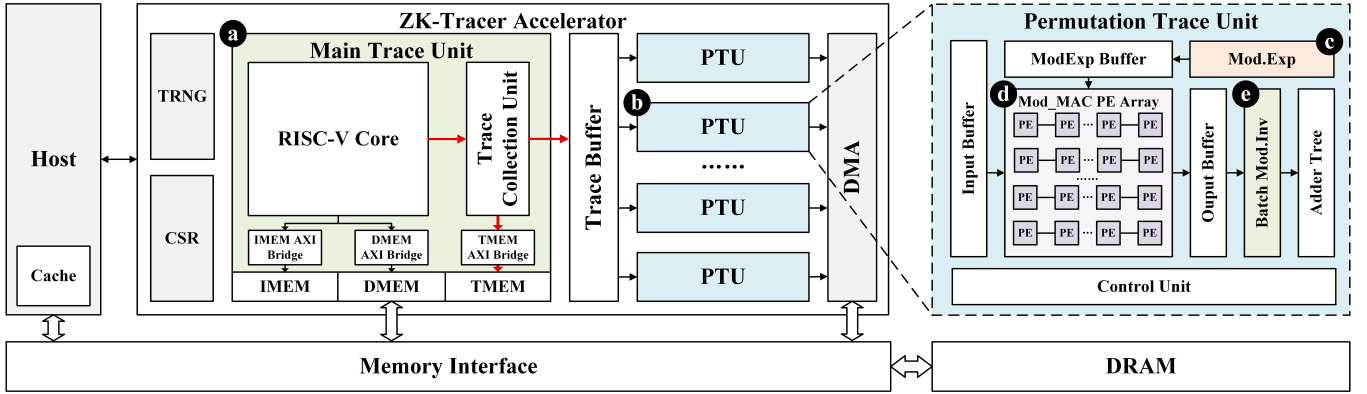}
  \caption{Architecture of ZK-Tracer}
  \label{fig:ZK-Tracer_Architecture}
  \vspace{-1.5em} 
\end{figure*}

\subsection{Architecture and Workflow}

Figure \ref{fig:ZK-Tracer_Architecture} illustrates the overall architecture of ZK-Tracer, our proposed dedicated hardware accelerator. ZK-Tracer is designed as a heterogeneous computing unit, tightly-coupled with a host CPU and sharing a unified physical memory space to enable efficient data exchange. The architecture is built around two core computation engines: the Main Trace Unit (MTU) and the parallel Permutation Trace Units (PTUs). The number of PTU instances is designed to match the number of trace tables required by the zkVM. These engines are supported by a suite of essential peripherals, including a True Random Number Generator (TRNG) for producing cryptographic challenges, a high-performance DMA engine, and a set of Control and Status Registers (CSRs) for host configuration and control. The MTU is responsible for executing the zkVM program to generate the Main Trace, while the parallel PTUs compute the Permutation Trace necessary to ensure cross-table consistency.

The operational workflow of ZK-Tracer begins with the host CPU allocating trace memory, configuring the accelerator via Control and Status Registers (CSRs), and issuing a start command to offload the task. Once activated, the Main Trace Unit (MTU) executes zkVM instructions(RISC-V) from its dedicated memories (IMEM\slash DMEM), generating the main trace on a row-by-row basis. In a pipelined fashion, each generated main trace row is concurrently written to main memory via the TMEM interface and simultaneously forwarded to an on-chip Trace Buffer. Working in parallel, the Parallel Trace Units (PTUs) consume trace rows from this buffer to compute the auxiliary columns, utilizing a DMA engine to write the results back to main memory. Finally, upon task completion, ZK-Tracer interrupts the host CPU, which can then access the complete trace data to proceed with the subsequent proof generation phase.

The on-chip Trace Buffer acts as a high-speed data channel between the MTU and PTU, establishing a deeply pipelined execution model. This design choice obviates the need for the PTU to repeatedly read vast amounts of main trace data from high-latency off-chip memory, thereby significantly reducing latency, memory bandwidth contention, and power consumption. Furthermore, dedicated memory interfaces for the MTU write-back path and the PTU DMA path ensure conflict-free parallel data transfers.

\subsection{Main Trace Unit (MTU)}

To achieve efficient and flexible hardware-based trace generation, we employ two key synergistic designs on a standard RISC-V core: microarchitectural extensions for non-intrusive trace capture, and Instruction Set Architecture (ISA) customizations for fine-grained, software-defined control.
The Main Trace Unit (MTU) is responsible for accurately executing the zkVM RISC-V instruction set and, in the process, synchronously generating the deterministic main execution trace. This trace constitutes the raw data required by the subsequent zero-knowledge proof system for proof generation.

\textbf{Core Design Choice.} We designed the MTU as a classic five-stage, in-order RISC-V core, a decision driven by two key requirements of Zero-Knowledge Proofs (ZKP): determinism and energy efficiency. The in-order execution model inherently generates the reproducible traces required for ZKP, thereby avoiding the complex hardware and verification overhead associated with instruction reordering in out-of-order (OoO) architectures. Furthermore, as a dedicated accelerator, its primary design goal is to maximize energy efficiency. A simple in-order core represents a more optimal choice compared to its power-hungry and costly OoO counterparts.

\begin{figure}[t]
  \centering
  \includegraphics[width=1\linewidth]{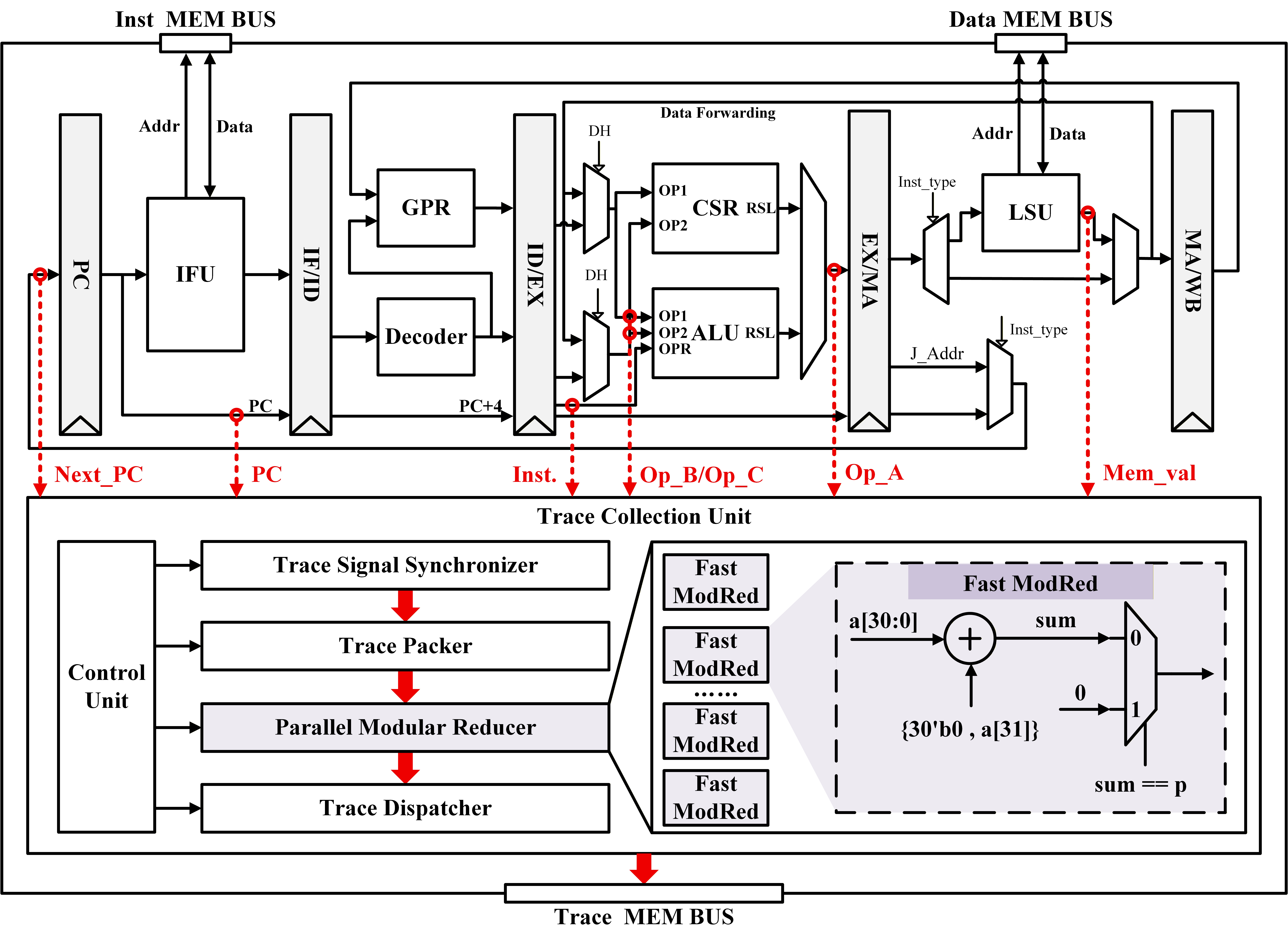}
  \caption{Main Trace Unit }
  \label{fig:MTU}
\end{figure}

\textbf{Microarchitectural Extensions.} To capture the execution trace without interfering with the core execution flow, we integrate a non-intrusive Trace Collection Unit (TCU) into the pipeline. As shown in Figure \ref{fig:MTU}, the TCU acts as a side-path module that directly snoops key pipeline paths. In the Execution (EX) and Memory (MEM) stages, the TCU concurrently captures a structured set of data, including the PC, operands, ALU result, and memory access values. The captured 32-bit raw datas are then immediately converted on-the-fly into Babybear field elements, as required by the proof system, by a bank of parallel, lightweight modular reduction units (FastModRed). This process, requiring only a single addition, has negligible computational overhead and introduces no back-pressure to the pipeline. Finally, the processed trace data are duplicated and dispatched to two destinations: one copy is written back to main memory via a dedicated trace bus, while the other is sent to an on-chip trace buffer.

\textbf{Extension of the Instruction Set Architecture (ISA).} Indiscriminate tracing of an entire program generates a massive volume of trace data, leading to prohibitive system overhead. To enable software-defined, on-demand trace generation, we extend the RISC-V ISA~\cite{wang2021xbgas,wang2020remote} with two custom instructions: \texttt{trace\_on} and \texttt{trace\_off}.
When the instruction decoder detects \texttt{trace\_on}, the TCU begins capturing the trace starting from the subsequent instruction. Conversely, upon detecting \texttt{trace\_off}, it suspends this activity. By inserting these instructions to bracket a target code segment, developers or compilers can precisely demarcate the region of interest for proving. This fine-grained control mechanism offers significant flexibility, drastically reducing unnecessary storage, bandwidth, and power consumption.

\subsection{Permutation Trace Unit (PTU)}

To enforce cross-table consistency, the LogUp protocol requires the generation of a Permutation Trace. This process is computationally intensive, involving a massive volume of modular arithmetic, which constitutes a significant performance bottleneck in the zkVM trace generation flow. To address this bottleneck, we designed a dedicated Permutation Trace Unit. As illustrated in Figure \ref{fig:ZK-Tracer_Architecture} \circled{b}, the PTU decomposes the entire computation into a four-stage, sequentially cascaded pipeline, with each stage realized by a specialized hardware module: a Modular Exponentiation Unit, a Systolic Array for Modular Multiply-Accumulate, a Batch Modular Inversion Unit, and a Parallel Prefix adder Tree.

\textbf{Modular Exponentiation Unit.} 
As depicted in Figure ~\ref{fig:ZK-Tracer_Architecture} \circled{c}, this unit is designed to calculate the modular exponentiation of the random challenge $\beta$.
Our design leverages a key structural property of zkVM traces: they are ``narrow and long'' ,where the number of columns, $k$, is typically small (from a few to hundreds), while the number of rows, $N$, can be in the millions or more.
We discard the inefficient method of recomputing weights $\{\beta^j\}$ for each row, instead adopting a precomputation and lookup strategy.
As a preprocessing step, the unit computes the entire set of required weights in a single batch using a square-and-multiply modular exponentiation algorithm with Montgomery optimization. These weights are then populated into a dedicated on-chip SRAM-based Look-Up Table (LUT).
This design amortizes the cost of expensive modular exponentiations from $O(N \cdot j)$ to $O(j)$ and removes it from the critical path of per-row processing, thereby providing low-latency weight access for the subsequent pipeline stages.

\begin{figure}[t]
  \centering
  \includegraphics[width=0.9\linewidth]{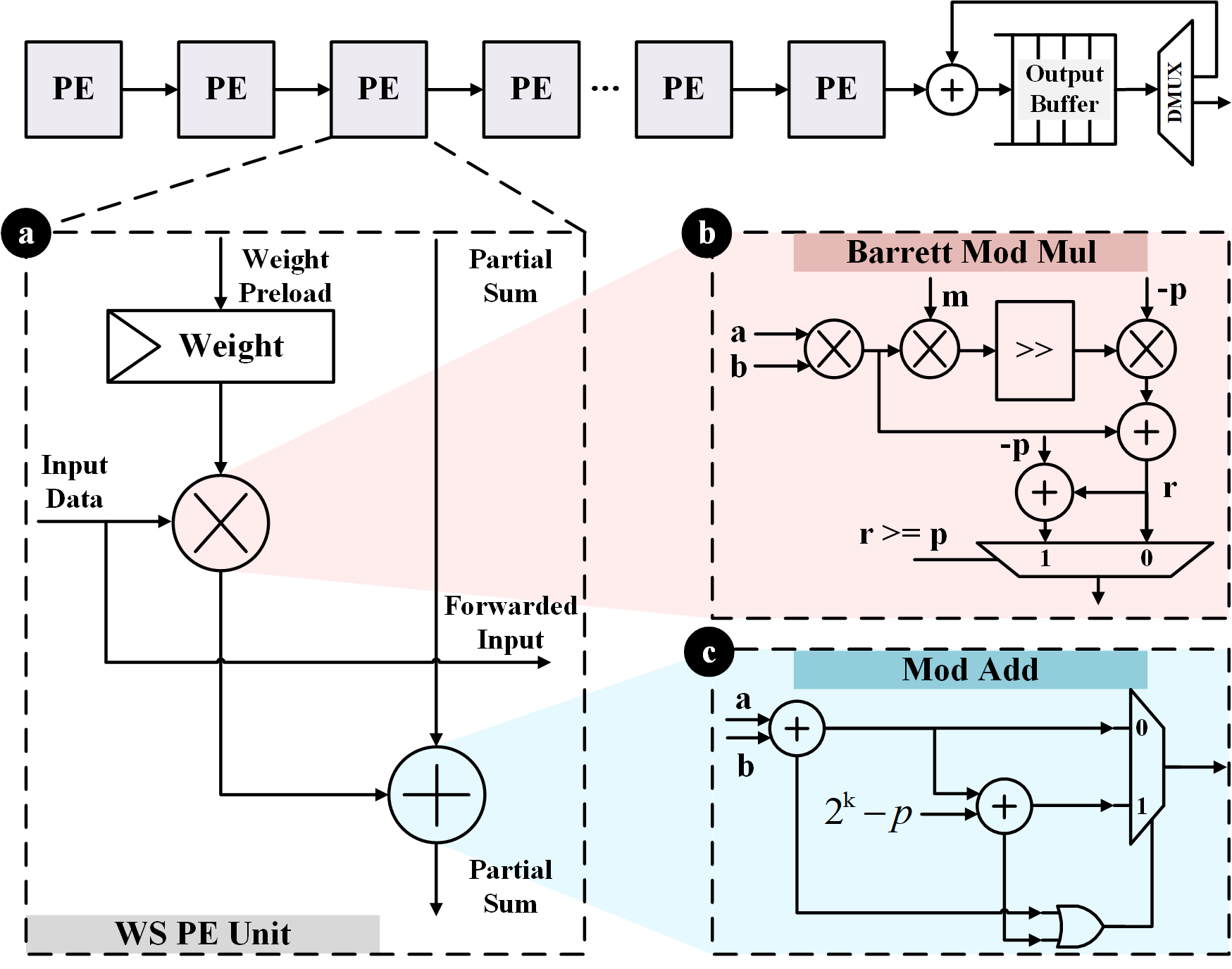}
  \caption{MMAC Systolic Array }
  \label{fig:pe_array}
\end{figure}

\begin{figure}[t]
  \centering
  \includegraphics[width=0.9\linewidth]{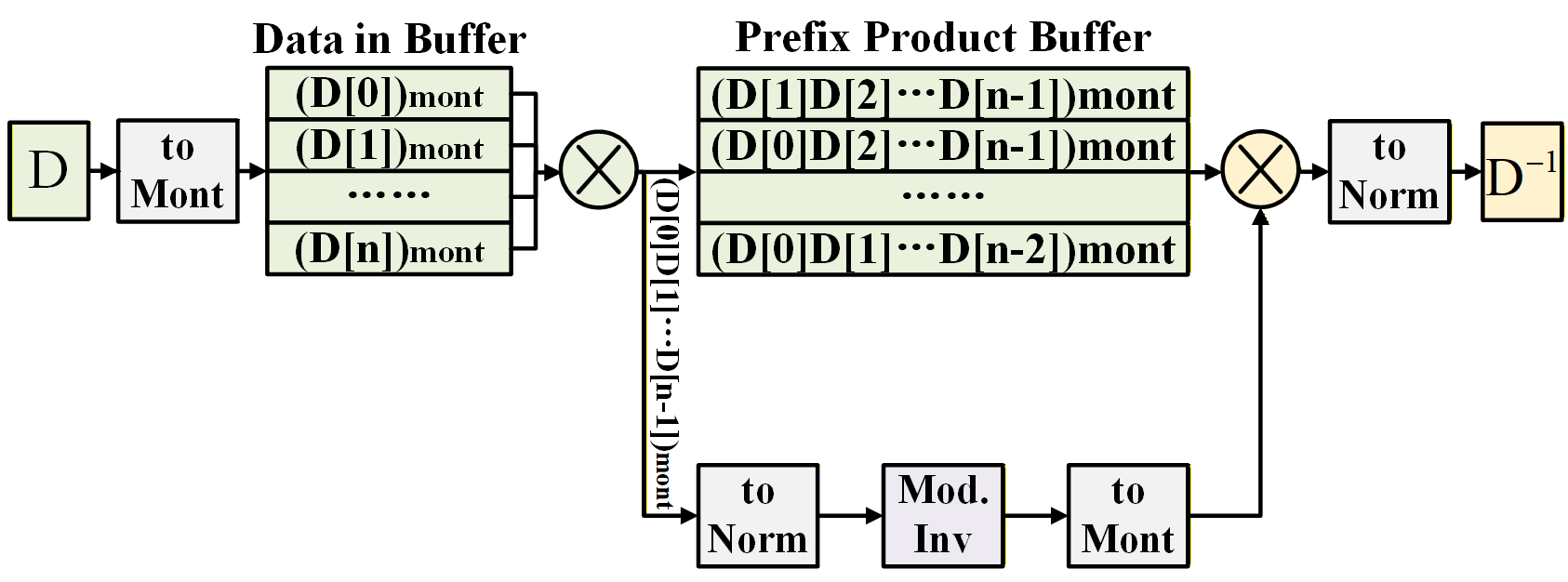}
  \caption{Batch Modular Inverse Unit }
  \label{fig:batched_inverse}
\end{figure}

\textbf{Modular Multiply-Accumulate (MMAC) Systolic Array.} 
As depicted in Figure ~\ref{fig:ZK-Tracer_Architecture} \circled{d}, this unit comprises a bank of parallel, weight-stationary, one-dimensional systolic arrays. This design enables concurrent processing of multiple trace rows, thereby dramatically boosting overall throughput. The circuit structure of the systolic array is depicted in Figure ~\ref{fig:pe_array}. The core arithmetic logic within its Processing Elements (PEs) is meticulously engineered to reduce resource consumption: modular multiplication employs the Barrett reduction algorithm to circumvent the domain conversion overhead associated with Montgomery methods, which is incompatible with streaming architectures; modular addition is realized via an optimized circuit that replaces a costly comparator with a single OR gate. During operation, pre-computed weights are loaded once and held stationary within all PEs, after which multiple rows of trace data are streamed into the arrays in parallel. For tables with smaller column widths, the control logic dynamically disables redundant PEs via clock gating to reduce power consumption. For wider tables, we designed an efficient partial sum feedback mechanism to complete the computation through multiple iterations, ensuring both architectural generality.

\textbf{Batch Modular Inverse Unit.}
As depicted in Figure ~\ref{fig:ZK-Tracer_Architecture} \circled{e}, this unit calculates the modular inverse. 
The generation of the permutation column requires modular inversions. 
However, computing these inverses individually, with a complexity of $O(N \log p)$, would become the new performance bottleneck. 
To address this, PTU integrates a Batch Modular Inverse Unit, as illustrated in Figure ~\ref{fig:batched_inverse}. 
This unit amortizes the computational cost of $N$ independent inversions to just a single modular inversion and approximately $3(N-1)$ modular multiplications. 
This reduces the overall complexity to nearly $O(N)$. 
The single required inversion is implemented using the Extended Euclidean Algorithm, thereby significantly enhancing the generation efficiency of the permutation column.

\section{Evaluation}
\subsection{Experimental Setup}

To the best of our knowledge, ZK-Tracer is \textbf{the first hardware accelerator} for zkVM Trace generation. 
To comprehensively evaluate its performance, we selected the industry-leading SP1 zkVM~\cite{SP1} as the performance baseline. SP1 was chosen because it is widely recognized as the fastest and most efficient open-source zkVM. 
We used the 8 benchmark programs officially provided by SP1 for testing, which cover a wide range of computation types and scales (recursive algorithms, cryptographic computations, and blockchain tasks). These benchmarks are diverse and representative, providing a comprehensive evaluation of ZK-Tracer performance. In terms of hardware implementation, ZK-Tracer uses the TSMC 28nm process for ASIC synthesis, with its performance, power, and area (PPA) metrics obtained from Synopsys Design Compiler. For DRAM power consumption evaluation, we integrated the DDR4 model from DRAMSim3~\cite{DRAMsim3}. Detailed experimental environment configurations are summarized in Table~\ref{tab:exp_setup}.

\begin{table}[!t]
\centering
\caption{Experimental Setup}
\vspace{-0.5em}
\label{tab:exp_setup}

\setlength{\tabcolsep}{4pt}
\renewcommand{\arraystretch}{0.9}

\begin{tabularx}{\columnwidth}{@{}l l X@{}}
\toprule
\textbf{Category} & \textbf{Component} & \textbf{Specification} \\
\midrule
CPU Baseline
& CPU & Intel Xeon E7-8860 v4 @ 2.20GHz, 64 cores \\
& Memory & 2 TB \\
& OS & Ubuntu 22.04 \\
& zkVM & Succinct SP1 v1.0.1 \\
& Compiler & rustc 1.87.0 \\
\midrule
ASIC Implementation
& HDL & SystemVerilog \\
& Process & TSMC 28nm HPC+ \\
& Synthesis Tool & Synopsys DC \\
\bottomrule
\end{tabularx}

\vspace{-0.8em}
\end{table}

\begin{figure*}[t!]
  \centering %
  \begin{minipage}[b]{0.26\linewidth}
    \centering
    \includegraphics[width=\linewidth, height=0.3\textheight, keepaspectratio]{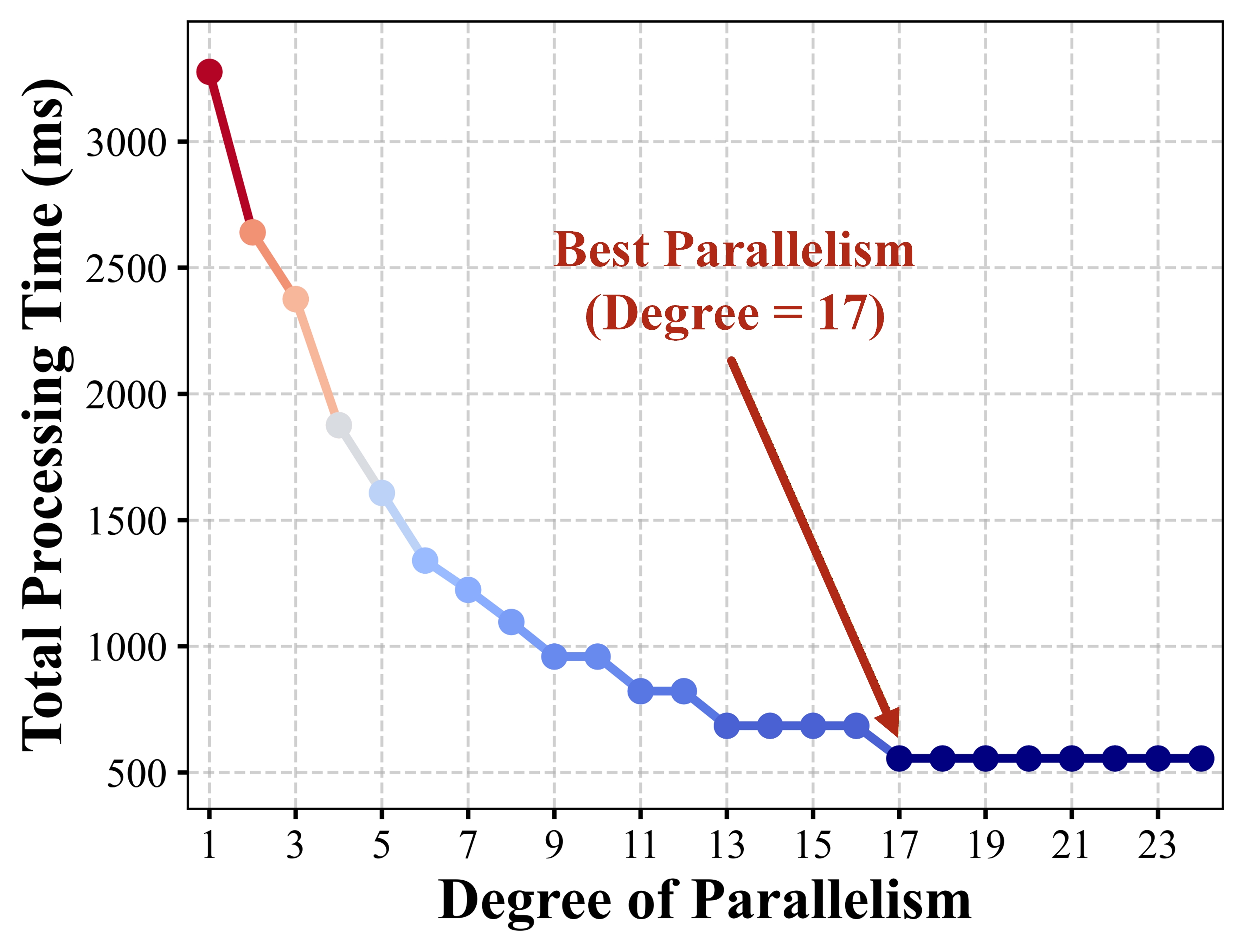}
    \captionof{figure}{Parallelism Analysis}
    \label{fig:dse-gem5}
  \end{minipage}
  \hfill
  \begin{minipage}[b]{0.4\linewidth}
    \centering
    \includegraphics[width=\linewidth, height=0.3\textheight, keepaspectratio]{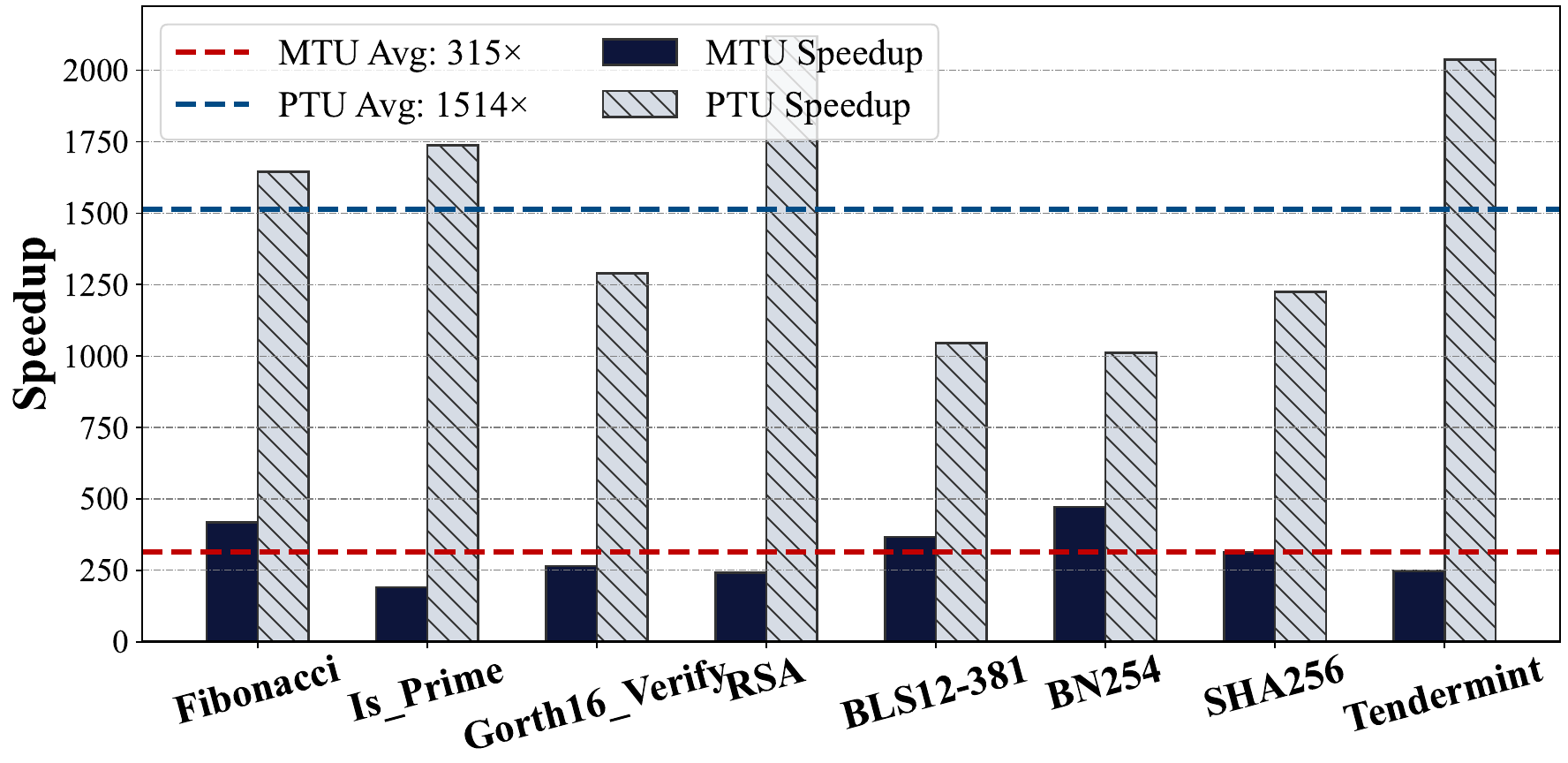}
    \captionof{figure}{MTU and PTU Speedup}
    \label{fig:MTUPTU_Speedup}
  \end{minipage}
  \hfill 
  \begin{minipage}[b]{0.3\linewidth}
    \centering
    \includegraphics[width=\linewidth, height=0.3\textheight, keepaspectratio]{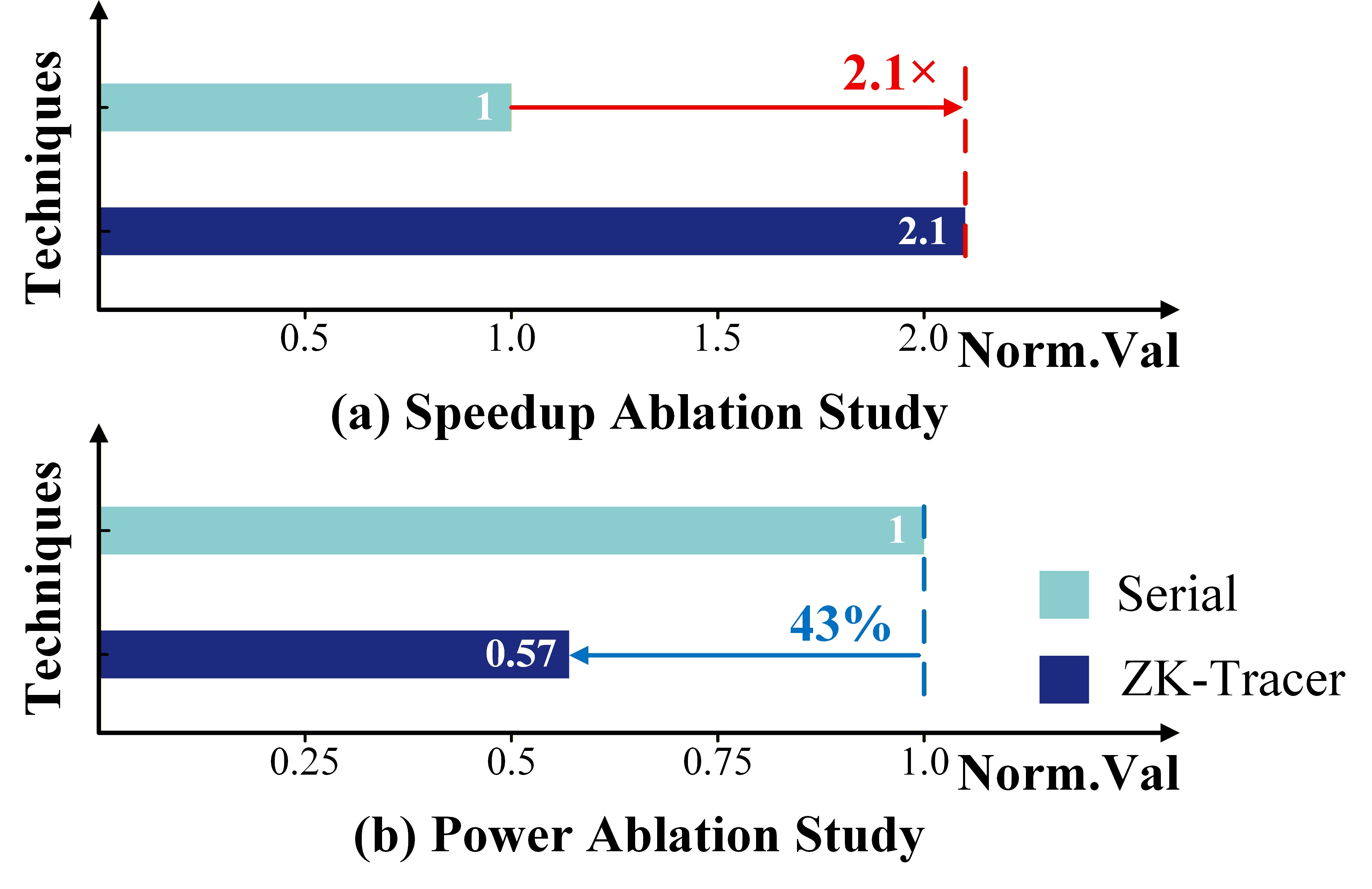}
    \captionof{figure}{Ablation Study}
    \label{fig:Ablation_Study}
  \end{minipage}
\end{figure*}

\subsection{Performance Model Simulation}

\textbf{Performance Overhead Analysis.} We first evaluate the performance overhead of ZK-Tracer in GEM5~\cite{gem5:software}. By comparing a standard MinorCPU core against a core integrated with the ZK-Tracer on the CoreMark benchmark suite, we measure a negligible IPC degradation of approximately \textbf{0\%} and no significant increase in memory access latency. This minimal overhead is attributed to our non-intrusive, snooping-based design, which completely decouples the tracing module from the processor's critical execution path. Consequently, it enables trace generation capabilities without compromising native performance.

\textbf{ZK-Tracer Architecture Parameter Analysis.} We leverage the GEM5 model to explore the design space for the ZK-Tracer PTU parallelism, balancing performance against cost. As shown in Figure~\ref{fig:dse-gem5}, the sweet spot is 17 parallel Compute Units (MMAC Array and Batch Mod.Inv), where the PTU throughput matches the MTU supply rate, maximizing hardware efficiency. We therefore select this cost-effective configuration for the final ASIC architecture.

\subsection{Hardware Implementation}

Following the insights and architectural parameters derived from our GEM5 C++ model, we implemented the proposed ZK-Tracer accelerator using SystemVerilog (approx. 27 kLoC) to evaluate its hardware overhead and performance. Our design is based on the open-source and silicon-verified SCR1 RISC-V core~\cite{SCR1}. The power, performance, and area (PPA) results of ZK-Tracer compared to the baseline are summarized in Table~\ref{tab:asic_ppa}. The results show that the functionality of the MTU was implemented by extending the SCR1 core, with only a \textbf{35\%} and \textbf{5\%} increase in area and power overhead, respectively. The complete ZK-Tracer accelerator operates reliably at \textbf{100~MHz} without timing violations. The total chip area is only \textbf{0.21~mm²}, and the power consumption is as low as \textbf{51.167~mW}.

\begin{table}[tbp]
\centering
\caption{ZK-Tracer PPA Results}
\label{tab:asic_ppa}

\setlength{\tabcolsep}{4pt}   

\begin{tabular}{lrr|>{\bfseries}r}
\toprule
\textbf{PPA} & \textbf{SCR1 (Baseline)} & \textbf{MTU (Overhead)} & \textbf{ZK-Tracer} \\
\midrule
Frequency & 100MHz & 100MHz (+0\%) & 100MHz \\
Area & 0.020mm$^2$ & 0.026mm$^2$ (+35\%) & 0.210mm$^2$ \\
Power & 0.140mW & 0.147mW (+5\%) & 51.167mW \\
\bottomrule
\end{tabular}

\vspace{-0.6em}  
\end{table}

\subsection{Performance}

Table~\ref{tab:performance} shows the comparison of execution time between ZK-Tracer and a CPU (running SP1). The experimental results show that ZK-Tracer demonstrates an overwhelming performance advantage in all tests, achieving an average speedup of \textbf{1829$\times$} over the pure software solution. This demonstrates the potential of a dedicated hardware architecture for accelerating core zkVM computational tasks. Further analysis (as shown in Figure~\ref{fig:MTUPTU_Speedup}) reveals that the MTU accelerated the program execution and Main Trace generation by \textbf{315$\times$}, while the PTU achieved a more significant speedup of \textbf{1514$\times$} by leveraging the high degree of parallelism in its computation.

\begin{table}[tbp]
  \centering
  \renewcommand{\arraystretch}{0.88}
  {\fontsize{9pt}{10pt}\selectfont
  \caption{Performance Comparison}
  \begin{tabularx}{\columnwidth}{Xrrr}
    \toprule
    \textbf{Benchmark} & \textbf{CPU (s)} & \textbf{ZK-Tracer (ms)} & \textbf{Speedup} \\
    \midrule
    Fibonacci    & 5.6      & 2.7            & 2063 $\times$ \\
    Is\_Prime    & 4.1      & 2.1            & 1929 $\times$ \\
    Gorth16\_Verify & 5.5      & 3.6            & 1554 $\times$ \\
    RSA          & 102.5    & 43.4           & 2362 $\times$ \\
    BLS12-381    & 477.6    & 338.0          & 1413 $\times$ \\
    BN254        & 379.1    & 255.4          & 1484 $\times$ \\
    SHA256       & 88.5     & 57.5           & 1538 $\times$ \\
    Tendermint   & 107.4    & 47.0           & 2286 $\times$ \\
    \midrule
    \textbf{Average} & -- & -- & \textbf{1829 $\times$} \\
    \bottomrule
  \end{tabularx}
  \label{tab:performance}
  }
  \vspace{-15pt}
\end{table}

\subsection{Ablation Analysis}

We conduct an ablation study to validate the benefits of our MTU and PTU pipeline design. 
Conventional approaches are constrained by an inefficient "write-back-then-read" main memory access pattern. 
In contrast, ZK-Tracer overcomes this bottleneck by introducing an on-chip Trace Buffer and increasing the parallelism of custom computation units to match the pipeline throughput. 
This enables deep pipelining of MTU trace generation and PTU trace computation, hiding main memory access latency. 
The experiment compares ZK-Tracer against an ablated version that reverts to the traditional sequential access mode. 

Figure~\ref{fig:Ablation_Study} show that our fine-grained pipeline design achieves \textbf{2.1$\times$ performance improvement} and \textbf{43\% reduction in power consumption}.

\section{Conclusion}

In this paper, we implemented ZK-Tracer, the first dedicated hardware accelerator to address the frontend execution and trace generation bottleneck in zkVMs. At its core lies a heterogeneous co-design architecture, featuring a Main Trace Unit and an Permutation Trace Unit. Our ASIC implementation of ZK-Tracer accelerates trace generation by up to 1829$\times$ compared to a high-performance CPU. This component-level speedup translates to a projected end-to-end acceleration of up to 963$\times$ when integrated with a state-of-the-art prover. 
ZK-Tracer fills a critical gap in the current landscape of ZKP hardware acceleration, marking a decisive step toward building full-stack, end-to-end high-performance ZKP systems.

\vspace{-6pt}
\begin{acks}
This work is supported by the National Natural Science Foundation of China under NSFC (Grant No. 92464301), the National Key Research and Development Program  (Grant No.2024YFB4405600), and the Key Research and Development Program of Jiangsu Province (Grant No.BG2024010).
\end{acks}

\bibliographystyle{ACM-Reference-Format}
\bibliography{references}

\end{document}